\documentclass[iop]{emulateapj}
\usepackage{amsmath}
\usepackage{color}

\shorttitle{}
\shortauthors{}
\begin{document}


\title{Eccentricity growth and orbit flip in near-coplanar hierarchical three body systems}
\author{Gongjie Li \altaffilmark{1}, Smadar Naoz \altaffilmark{1}, Bence Kocsis \altaffilmark{1, 2}, Abraham Loeb \altaffilmark{1}}
\affil{$^1$ Harvard-Smithsonian Center for Astrophysics, The Institute for Theory and
Computation, \\60 Garden Street, Cambridge, MA 02138, USA \\
$^2$ Institute for Advanced Study, Princeton, NJ 08540, USA}
\email{gli@cfa.harvard.edu}


%\altaffiltext{1}{Harvard-Smithsonian Center for Astrophysics, The Institute for Theory and
%Computation, 60 Garden Street, Cambridge, MA 02138, USA}
%\altaffiltext{2}{Institute for Advanced Study, Princeton, NJ 08540, USA}
%\altaffiltext{2}{Society of Fellows, Harvard University.}
%\altaffiltext{3}{present address: Center for Astrophysics,
%   60 Garden Street, Cambridge, MA 02138}
%\altaffiltext{4}{Visiting Programmer, Space Telescope Science Institute}
%\altaffiltext{5}{Patron, Alonso's Bar and Grill}

\begin{abstract}
The secular dynamical evolution of a hierarchical three body system, in which a distant third object orbits around a binary has been studied extensively, demonstrating that the inner orbit can undergo large eccentricity and inclination oscillations. It was shown before that starting with a circular inner orbit, large mutual inclination ($40^\circ - 140^\circ$) can produce long timescale modulations that drive the eccentricity to extremely large value and can flip the orbit. Here, we demonstrate that starting with an almost coplanar configuration, for eccentric inner and outer orbits, the eccentricity of the inner orbit can still be excited to high values, and the orbit can flip by $\sim180^\circ$, rolling over its major axis. The $\sim180^\circ$ flip criterion and the flip timescale are described by simple analytic expressions that depend on the initial orbital parameters. With tidal dissipation, this mechanism can produce counter-orbiting exo-planetary systems. In addition, we also show that this mechanism has the potential to enhance the tidal disruption or collision rates for different systems. Furthermore, we explore the entire $e_1$ and $i_0$ parameters space that can produce flips.
\bigskip
\end{abstract}

%\keywords{Earth--chaos--instabilities, Exoplanets--dynamics}

\section{Introduction}
The Kozai-Lidov mechanism \citep{Kozai62, Lidov62} has proven very useful for interpreting different astrophysical systems. For example, it has been shown that its application can explain Hot Jupiters configurations and obliquity \citep[e.g.][]{Holman97, Wu03, Fabrycky07, Veras10, Correia11, Naoz11N, Naoz12}. Furthermore, close stellar binaries with two compact objects are likely produced through triple evolution, and secular effects may play key role in these systems and in their remnants (e.g. \citealt{Harrington69, Mazeh79, Soderhjelm82, Kiseleva98, Ford00, Eggleton01, Fabrycky07, Perets09F, Thompson11, Katz12, Shappee13, Naoz13, Naoz14}). Secular effects have been proposed as an important element both in the growth of black holes at the centre of dense star clusters and the formation of short-period binaries black hole \citep{Blaes02, Miller02, Wen03} and tidal disruption events \citep{Chen09, Chen11, Wegg11, Bode13, Li13}.

The Kozai-Lidov mechanism was first discussed by \citet{Kozai62} and \citet{Lidov62}, who applied the mechanism for specific configurations where the outer orbit was circular and one of the members inner binary was a test (massless) particle. In this situation, the component of the inner orbit's angular momentum projected on the total angular momentum of the whole system (z axis) is conserved. To lowest order, the quadrupole approximation provides a valid presentation of the system \citep{Lidov74}. In that case, the system is integrable and the eccentricity and the inclination undergo large oscillations when $i>39.2$ degree due to the ``Kozai resonance" \citep{Thomas96}.

Recently, \cite{Naoz11N,Naoz12} showed that relaxing either one of these assumptions, (i.e., an eccentric outer orbit, or non-negligible mass binary members) leads to qualitatively different behavior. In this case the z-component of the inner, and outer orbit's angular momentum is not conserved.  Considering systems beyond the test particle approximation, or a circular orbit, requires the octupole--level of approximation \citep{Harrington68, Harrington69, Ford00, Blaes02}. 

The octupole approximation can lead to extremely large values for the inner orbit's eccentricity \citep{Ford00, Naoz13,  Teyssandier13}.  Furthermore, the inner orbit's inclination can flip its orientation from prograde to retrograde, with respect to the total angular momentum \citep{Naoz11N,Naoz13}.  We refer to this process as the {\it eccentric Kozai--Lidov} (EKL) mechanism. It has been shown in \citet{Naoz13} that the secular approximation can be used as a tool for understanding different astrophysical settings, from massive or stellar compact objects to planetary systems. 
%
%considered the secular evolution of  particles under the perturbation of an object on a circular orbit. In this case, the z component of the angular momentum of the particles are constant and the system is integrable. In addition, the eccentricity and the inclination undergo large oscillations when $i>39.2$ degree due to the Kozai resonance \citep{Thomas96}. 
%
%The recent development of this mechanism have shown interesting results. For instance, \citet{Naoz11N} noticed that the z component of the angular momentum of the inner orbit is not conserved if the inner object is not a test particle. In addition, when the outer orbit is non-circular, the z component of the angular momentum of the inner orbit is also not conserved. In this case, the octupole terms need to be taken into account. In particular, the eccentricity of the inner orbit may become very close to unity and produce tidal disruption events. Moreover, the inner orbit may flip from prograde to retrograde or vice versa \citep{Lithwick11, Katz11, Naoz13}. 
%

We focus on the octupole order when the inclination is set to be almost coplanar. \citet{Lee03} considered the case when the mutual inclination is zero, and they showed that the eccentricity can oscillate due to the octupole effects. Here we set the mutual inclination to be non-zero but still very small. We show both numerically and analytically, that an eccentric inner orbit ($e_1 > 0.6$) in almost coplanar configuration with an eccentric outer orbit becomes highly eccentric ($e_1\gtrsim 0.9999$) due to the octupole effects. Provided that it avoids a direct collision with or tidal disruption by the central object, it undergoes a $\sim180^\circ$ flip. We derive the flip criterion analytically (equation (\ref{eqn:criterion})), and we apply this mechanism to the retrograde hot jupiters and discuss its application to tidal disruptions. 

The paper is organized as follows. In \textsection{2}, we demonstrate the coplanar flip, and derive the analytical expression for the flip criterion and timescale. In \textsection{3}, we start the system with a large range of parameter space to study the flip criterion and timescale. Finally, in \textsection{4}, we discuss the applications of the coplanar flip to exo-planetary systems and tidal disruption events.

\section{Coplanar Flip}
\begin{figure}[h]
\begin{center}
\includegraphics[width=3.3in, height=2.5in]{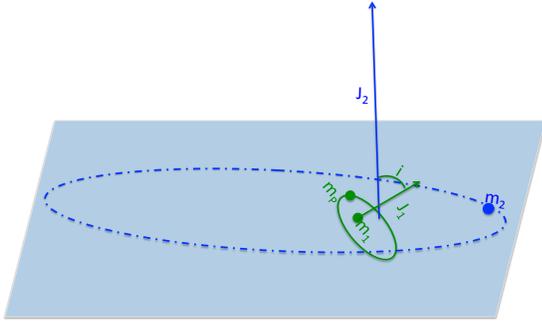} \label{fig:config}
\caption{{\it Configuration of the hierarchical 3-body system}.  An object $m_P$ orbits around the object $m_1$ and forms an inner binary. The outer binary is composed of the outer object $m_2$ orbiting the center mass of $m_1$ and $m_P$. The parameters of the inner and outer binary are denoted by subscripts 1 and 2, respectively. The angle $i$ represents the mutual inclination between the two orbits, and $J_1$ and $J_2$ represent the orbital angular momenta of the inner and outer binary. The near-coplanar case corresponds to $i \sim 0^\circ$.}
\end{center}
\end{figure}

The Kozai-Lidov mechanism relates to the hierarchical three-body system as shown in Figure \ref{fig:config}. The parameter $\epsilon$, 
\begin{equation}
\epsilon = \frac{a_1}{a_2}\frac{e_2}{1-e_2^2} ,
\end{equation}
 is small, where $a$ is the semi-major axis and $e$ is the eccentricity of the inner ``1" and outer ``2" orbit \citep{Naoz13}.

In the test particle quadrupole approximation ($m_P \to 0$, $e_2 = 0$), the Kozai-Lidov resonance is between the longitude of periapsis and the longitude of ascending node of the inner orbit \citep{Kozai62}. The eccentricity and the inclination oscillate with large amplitudes when the inclination is over 40 degree. This resonance also exists if the test particle mass is significant. The quadrupole approximation describes the orbital evolution when the outer orbit is circular. When the outer orbit is non-circular, the octupole approximation is needed, inducing variations in eccentricity and inclination on longer timescales, and causes excursions to even higher eccentricities and inclinations above $90^{\circ}$ \citep{Naoz11N, Naoz13}. However, starting with a circular inner orbit, the inclinations that produce this behavior are restricted to the range of $\sim 40^{\circ}-140^{\circ}$. 

\begin{figure}[h]
\begin{center}
\includegraphics[width=3.3in, height=2.5in]{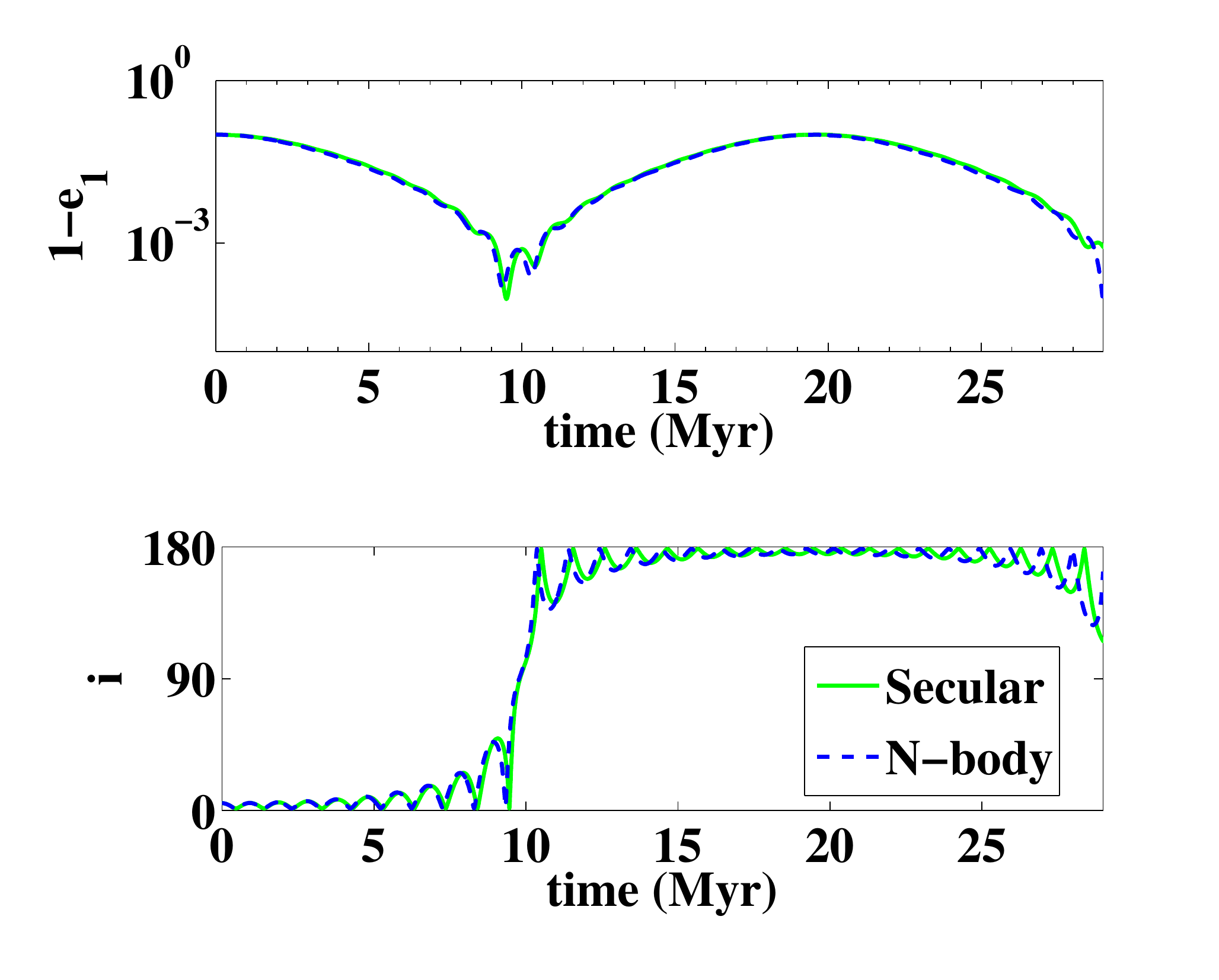}
\caption{{\it The consistency and convergence of the numerical method for the point mass dynamical evolution of the inner orbit}.  We set $m_1=1M_{\odot}$, $m_2=0.02M_{\odot}$, $m_P=10^{-3}M_{\odot}$, $a_1=1$ AU, $a_2=50$ AU, $i = 5^{\circ}$, $e_1=0.9$, $e_2=0.7$, $\omega_1=\omega_2=\Omega_2=0^{\circ}$ and $\Omega_1=180^{\circ}$. The green line represents the run integrated using the secular approximation, and the dashed blue line represents the results of the N-body simulation using the Mercury code. The results of the two methods agree. In both cases, the test particle exhibits an $180^{\circ}$ flip in a coplanar configuration.} \label{fig:directInt} 
\end{center}
\end{figure}

Starting with an almost coplanar configuration ($e_1 = 0.8$, $i = 5^\circ$), we find that the inner orbit can still flip if it starts eccentric (the high eccentricity low inclination case: hereafter HeLi). We show the flip in Figure \ref{fig:directInt} using direct N-body integrations, with the MERCURY software package \citep{Chambers97}. The remarkable agreement with the integration using the secular approximation up to the octupole order is also shown in Figure \ref{fig:directInt}.

\begin{figure}[h]
\begin{center}
\includegraphics[width=1.45in, height=2in]{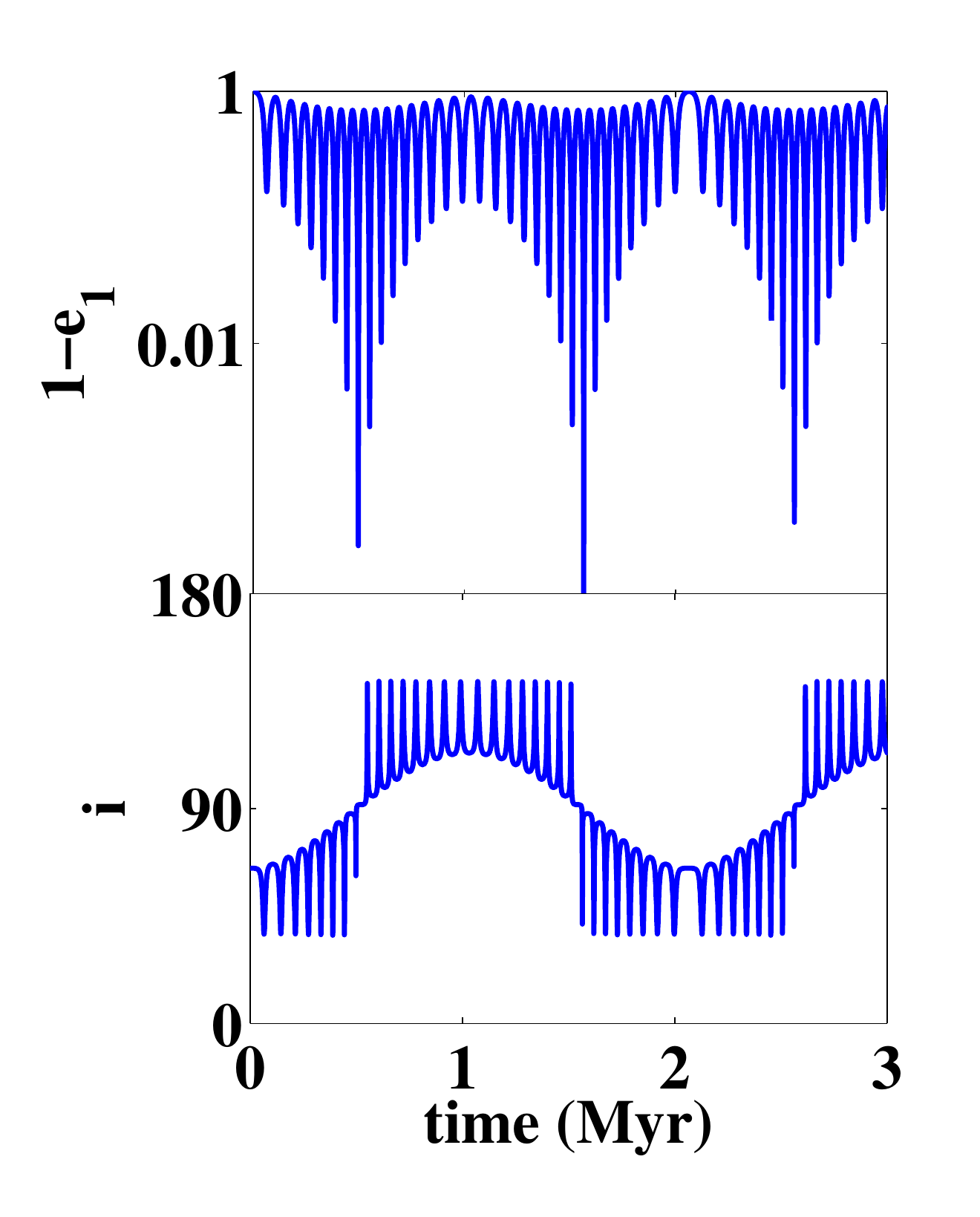}
\includegraphics[width=1.45in, height=2in]{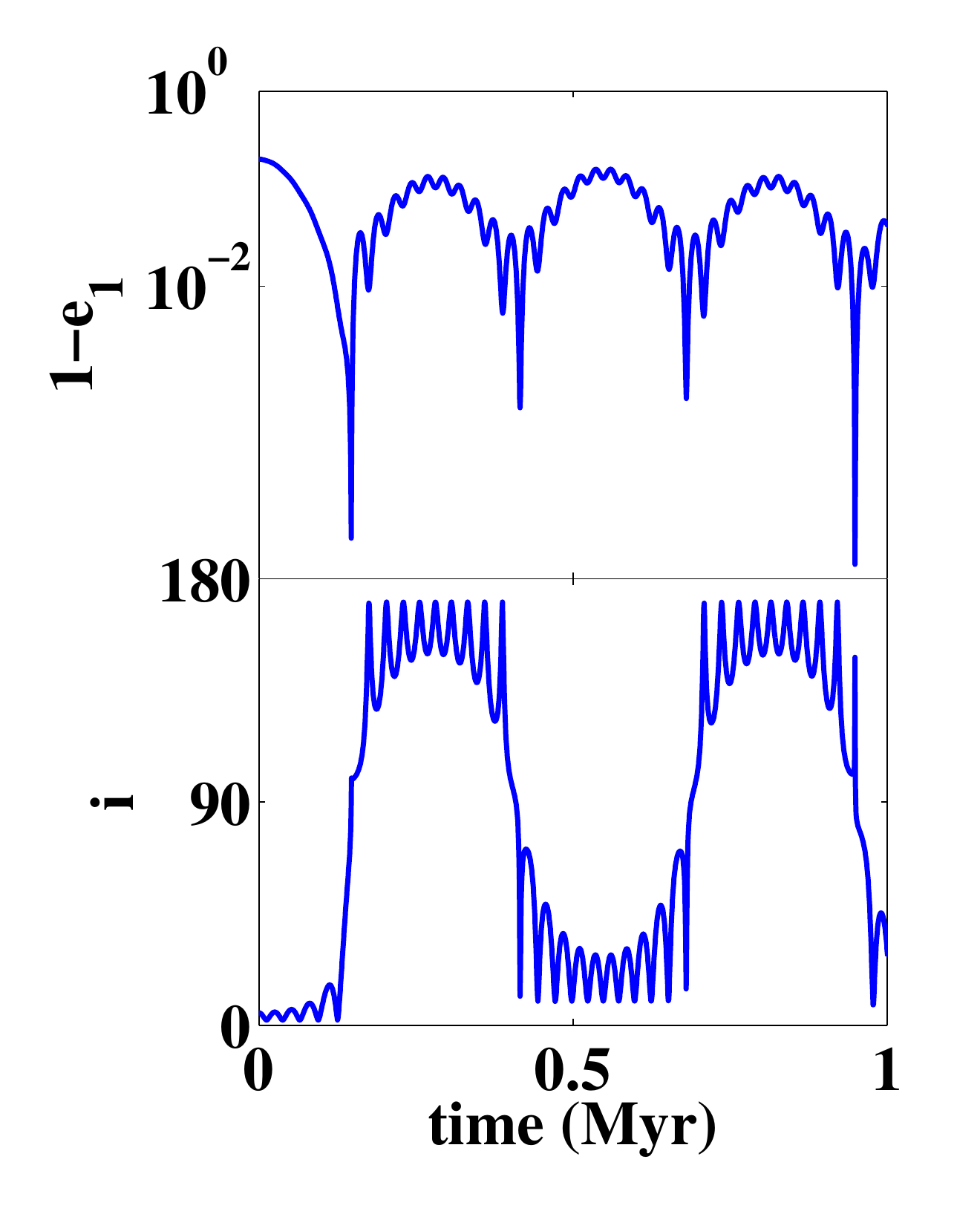}
\caption{{\it The evolution of the inner orbit's eccentricity and mutual inclination}.  We set the mass of $m_1$ and $m_P$ to a solar and a Jupiter mass, and the mass of the outer perturber $m_2$ to $0.03 M_{\odot}$, and $\omega_1 = 0^{\circ}$, $\Omega_1 = 180^{\circ}$, $e_2=0.6$, $a_1 = 4$ AU, $a_2 = 50$ AU. We use the secular approximation to calculate the dynamical evolution of point masses. The left panel shows the standard Kozai cycles for comparison, ($e_1 = 0.01$, $i = 65^{\circ}$), and the right panel shows the eccentric coplanar scenario ($e_1 = 0.8$, $i = 5^{\circ}$). For the former, both $i$ and $e_1$ oscillate with large amplitudes, but in the eccentric coplanar case, $e_1$ increases steadily and $i$ oscillates to maintain a coplanar configuration. The flip occurs much more rapidly in the eccentric coplanar case.} \label{fig:2ab} 
\end{center}
\end{figure}

The flip in the HeLi case is qualitatively different from the low eccentricity high inclination case (LeHi case, see Figure \ref{fig:2ab} left panel). Specifically, in the initially coplanar case, the oscillation amplitude of the inclination is small maintaining a coplanar configuration before the flip, as the eccentricity grows monotonically to large values. The timescale for the inclination to cross over $90^{\circ}$ (namely the flip timescale) is much shorter. Moreover, the underlying resonances responsible for the flips are different (Li et al. in prep.). The HeLi case is dominated by only octupole order resonances. However, the LeHi case is dominated by both the quadrupole order resonances and the octupole order resonances. As a comparison, we illustrate the difference in the HeLi case in the right panel of Figure \ref{fig:2ab}.

To illustrate the orbital evolutions, we show the movies\footnote{\url{https://www.cfa.harvard.edu/\~gli/images/lowi.mp4}; \\ \url{https://www.cfa.harvard.edu/\~gli/images/highi.mp4}} of the inner orbital evolution in the test-particle limit for both cases. We set the z axis to be aligned with the total angular momentum and the x axis is aligned with the ascending node of the outer orbit. In the test particle limit, the outer orbit is stationary. In the movies, the inner orbit is painted according to the value of the mean anomaly. The black arrow represents the normalized orbital angular momentum, and the pink arrow represents the z component of the angular momentum. The orbital flip can be observed in the rapid reorientation of the pink arrow from the $+z$ to the $Ðz$ direction. The black arrow shows the orientation of the orbit. The orbit rolls over its major axis when it flips. This can be understood analytically as $dJ_1/dt$ is perpendicular to the eccentricity vector at $i = 90^\circ$.

%, and the obliquity angle is restricted to roughly that range as well \citep{Thomas96}. 

%
%
%In this article, we focus on the case where the initial eccentricity is high and the initial inclination is almost coplanar. We find that the initially eccentric coplanar case is qualitatively different from the standard Kozai effect (figure \ref{fig:2ab}). Specifically, in the initially coplanar case, the oscillation amplitude of the inclination is small maintaining a coplanar configuration before the flip, as the eccentricity grows monotonically to large values. The timescale for the inclination to cross over $90^{\circ}$ (flip timescale) is much shorter. We demonstrate this process with N-body integration and explain the results with the octupole approximation. The remarkable agreement with the N-body integration is shown in figure \ref{fig:directInt}. 

\subsection{Analytical Derivation}

The coplanar flip phenomenon can be understood analytically in the test particle approximation (i.e, $m_P \to 0$). In the large inclination regime, it was shown that the behavior associated with the test particle approximation is valid for $m_2/m_P>7$ \citep{Teyssandier13}. 

This test particle approximation in hierarchical 3-body systems was studied extensively in the past \citep{Lithwick11, Katz11}, but only in the regime of large inclinations between the inner and outer orbit's \citep[and for small initial inner eccentricity $e_1<0.5$][]{Lithwick11}.  Our initial coplanar configuration simplifies the analytic treatment. The $\sim180^{\circ}$ flip occurs due to octupole-level terms, whose importance can be estimated via $\epsilon$. 

We follow the equation of motion using a Hamiltonian description for the non-relativistic hierarchical three body problem. We define the energy function as the negative of the Hamiltonian in the secular approximation up to the octupole level \citep{Lithwick11}. The Hamiltonian of such systems is well documented in the literature (e.g. \citet{Harrington68, Harrington69, Ford00}). The scaled energy function for the hierarchical three-body system in the test particle approximation to this order is $F_{quad} + \epsilon F_{oct}$:

\begin{align}
F_{quad} &= -(e_1^2/2)+\theta^2+3/2e_1^2\theta^2 \\
					&+ 5/2e_1^2(1-\theta^2)\cos(2\omega_1), \nonumber \\ 
F_{oct} &= \frac{5}{16}(e_1+(3e_1^3)/4) \\
			&\times ((1-11\theta-5\theta^2+15\theta^3)\cos(\omega_1-\Omega_1) \nonumber \\
			&+ (1+11\theta-5\theta^2-15\theta^3)\cos(\omega_1+\Omega_1)) \nonumber \\
			&-\frac{175}{64} e_1^3((1-\theta-\theta^2+\theta^3)\cos(3\omega_1-\Omega_1) \nonumber \\
			&+(1+\theta-\theta^2-\theta^3)\cos(3\omega_1+\Omega_1)),  \nonumber
\end{align}
where $\theta = \cos{i}$, $\omega_1$ is the argument of periapsis of the inner orbit and $\Omega_1$ is the longitude of ascending node of the inner orbit.

To the first order in $i$, the evolution of $e_1$ and $\varpi_1=\omega_1+\Omega_1$ can be solved (we denote $\varpi_1=\omega_1+\Omega_1$ hereafter). Specifically, $\dot{e_1}$ and $\dot{\varpi_1}$ depend only on $e_1$ and $\varpi_1$:
\begin{align}
\label{eqn:S3}
\dot{e_1} &=\frac{5}{8}J_1(3J_1^2-7)\epsilon\sin(\varpi_1), \\
\dot{\varpi_1} &= J_1\Big(2+\frac{5(9J_1^2-13)\epsilon\cos(\varpi_1)}{\sqrt{1-J_1^2}}\Big),
\end{align}
where $J_1=\sqrt{1-e_1^2}$. Combining the two differential equations, we can express $\cos{\varpi_1}$ as a function of $e_1$:
\begin{align}
\cos{\varpi_1} = \frac{8e_1^2-C}{e_1(20+15e_1^2)\epsilon},
\label{eqn:S5}
\end{align}
where $C$ is an integration constant, which is the energy that corresponds to $i=0$ and can be determined from the initial condition. Substituting $\cos(\varpi_1)$ in the differential equation of $\dot{e_1}$, we obtain a separable first order differential equation:
\begin{align}
\label{eqn:dedt}
\dot{e_1} = -\frac{5}{8}(4+3e_1^2)\sqrt{(1-e_1^2)\Big(1-\frac{(C-8e_1^2)^2}{25e_1^2(4+3e_1^2)^2\epsilon^2}\Big)\epsilon}.
\end{align}
Integrating equation (\ref{eqn:dedt}), we get $e_1$ as a function of time.

\begin{figure}[h]
\begin{center}
\includegraphics[width=1.4in, height=2in]{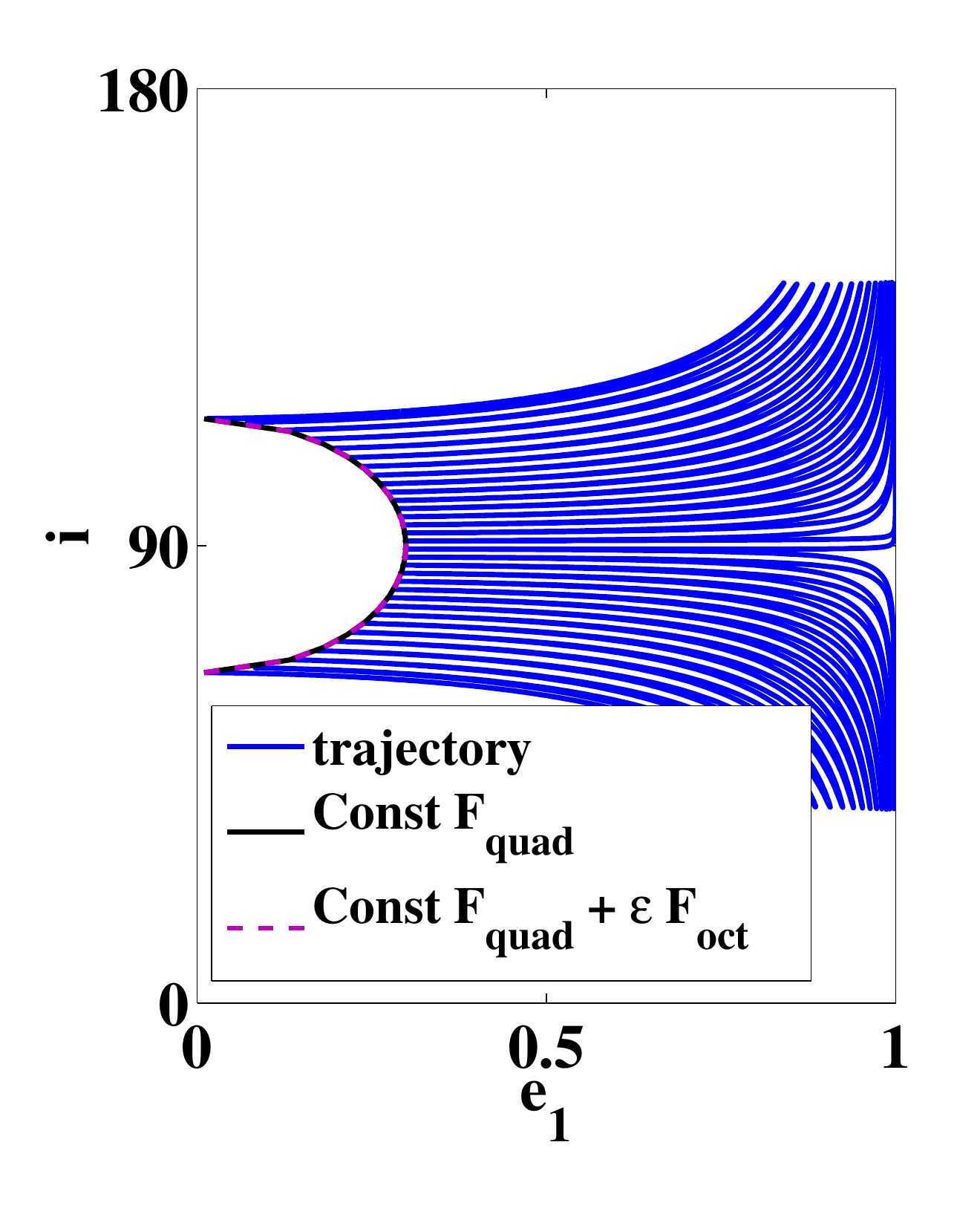}
\includegraphics[width=1.4in, height=2in]{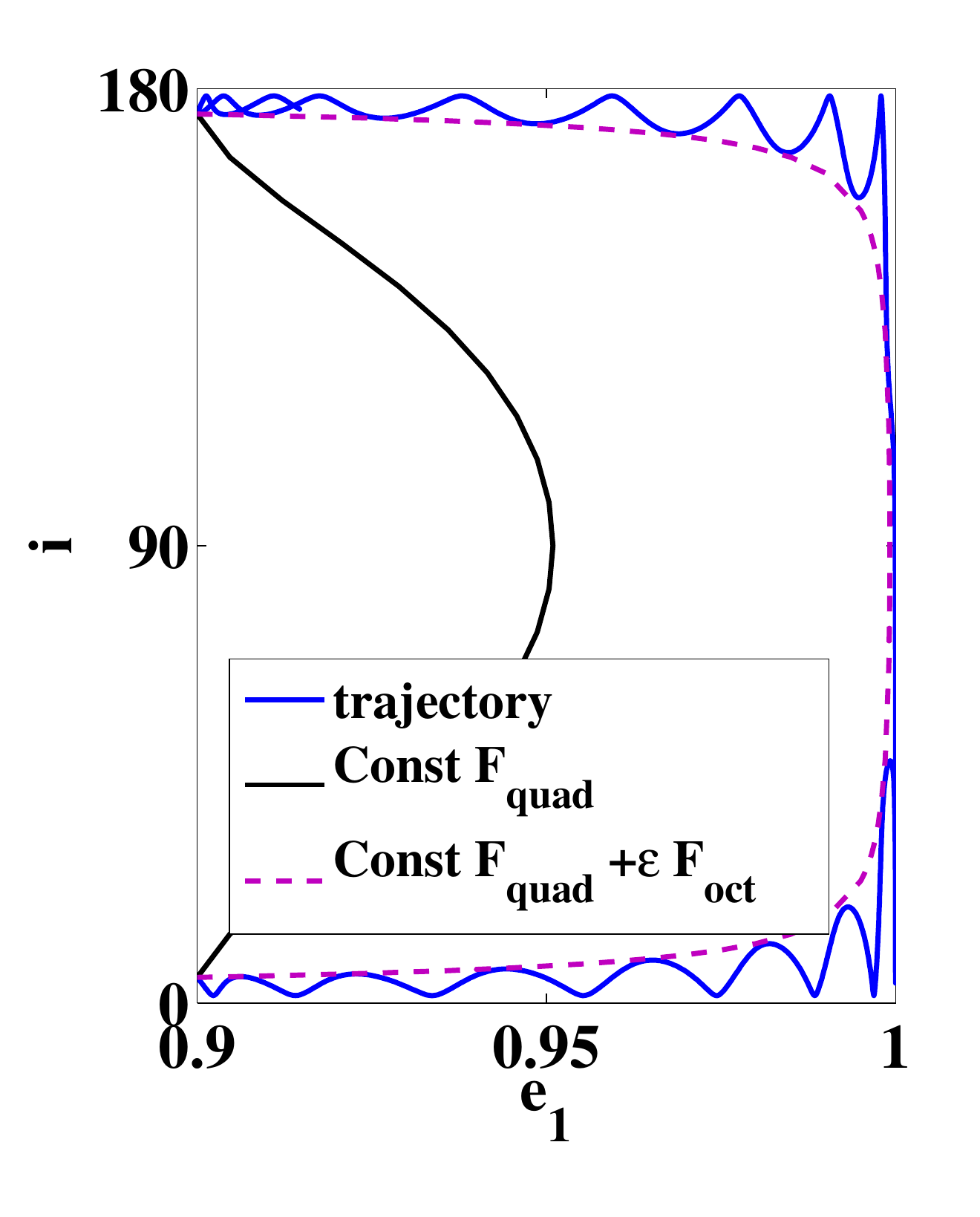}
\caption{Left Panel: standard Kozai-Lidov scenario with initial conditions $e_1=0.01$, $i = 65^{\circ}$, $m_1 = 0.3M_{\odot}$, $m_2 = 0.1M_{\odot}$,  $a_1=1$ AU, $a_2=40$ AU, $\omega_1=0^{\circ}$, $\Omega_1=180^{\circ}$. Right Panel: the eccentric coplanar case, with initial conditions $e_1=0.9$, $i = 5^{\circ}$, $m_1=0.3M_{\odot}$, $m_2=0.03M_{\odot}$, $a_1=1$ AU, $a_2=40$ AU, $\omega_1=0^{\circ}$, $\Omega_1=180^{\circ}$. The evolution tracks represent the change of $J_z$ \citep{Lithwick11}. The inclination $i$ and $e_1$ oscillate for large initial inclinations, while in the low inclination case, $i$ oscillates and $e_1$ increases steadily. The dashed line represents the constant $F_{quad} +\epsilon F_{oct}$ curve at $\omega_1=0^{\circ}$, which sets the maximum or minimum inclination during a quadrupole cycle. The black solid line represents the constant $F_{quad}$ curve. The maximum inclination in each quadrupole Kozai cycle follows the constant $F_{quad}$  curve only in the HiLe mechanism.} \label{fig:S4} 
\end{center}
\end{figure}

%This analytic expression helps understand the evolution of the inner orbit.
Figure \ref{fig:2ab} shows that the eccentricity increases steadily and the inclination oscillates in the low inclination scenario until the flip occurs. This behavior can also be seen in Figure \ref{fig:S4}. The steady change of $e_1$ can be explained by equation (\ref{eqn:S3}). Since 
\begin{align}
\frac{5}{8}J_1(3J_1^2-7)\epsilon < 0 , 
\end{align}
as $(0<J_1<1)$, the sign of $\dot{e_1}$ depends on $\sin(\varpi_1)$, and $e_1$ reaches its extremum when $\sin(\varpi_1) = 0$. In addition, since $\varpi_1$ vanishes to the quadrupole order, the change of $\varpi_1$ is small. Thus, $e_1$ does not oscillate over the quadrupole timescale. Instead, $e_1$ increases or decreases monotonically to $e_{min}$ or $e_{max}$.

Using the conservation of $F_{quad} +\epsilon F_{oct}$, we can estimate the evolution of the inner orbit in the low inclination case by calculating the constant energy curve in Figure \ref{fig:S4} (pink dashed line). The total energy $F_{quad} + \epsilon F_{oct}$ depends on the four variables: $e_1$, $i$, $\omega_1$ and $\Omega_1$. To obtain the maximum inclination, $i_{max}$ as a function of $e_1$ as shown in Figure \ref{fig:S4}, we need to express $\omega_1$ and $\Omega_1$ as a function of $e_1$ at $i =i_{max}$. From the equation of motion, $\dot{i}\propto \sin(2\omega_1)$, thus the maximum of inclination occurs at $\omega_1=0$. When $\omega_1=0$, $\cos{\varpi}=\cos{\Omega}$, thus, substituting equation \ref{eqn:S5} in the conservation of $F_{quad} +\epsilon F_{oct}$, we get $i_{max}$ as a function of $e_1$. The analytic expression is compared with the numerical trajectory in Figure \ref{fig:S4}, where the evolution of $e_1$ and $i$ are obtained by integrating the equations of motion in the secular approximation. 

Moreover, Figure \ref{fig:S4} shows another major difference between the LeHi behavior and the HeLi case studied here. For the LeHi case, energy conservation of the quadrupole approximation, $F_{quad}$, can be used to find the maximum eccentricity and the minimum inclination. However, the octupole correction is non-negligible in the HeLi case. 

The flip time can be estimated using equation (\ref{eqn:dedt}). Since $\sin{\varpi_1} < 1$, $e_1$ increases steadily before the flip, the flip time scale can be estimated as: 
\begin{equation} 
t_{flip} = \int_{e_{min}}^{e_{max}} \dot{e_1}^{-1}~de \  .
\end{equation}. 

The initial conditions of this configuration are  $i \sim 0$, $e_{1,0} \to1$, where the subscript ``0" represents the initial condition. Since $e_1$ increases monotonically until the flip, we set the minimum eccentricity to be the initial eccentricity, i.e., $e_{min} = e_{1,0}$. Furthermore, the maximum eccentricity is simply $e_{max} = 1$. 

\begin{figure}[h]
\begin{center}
\includegraphics[width=3in, height=2in]{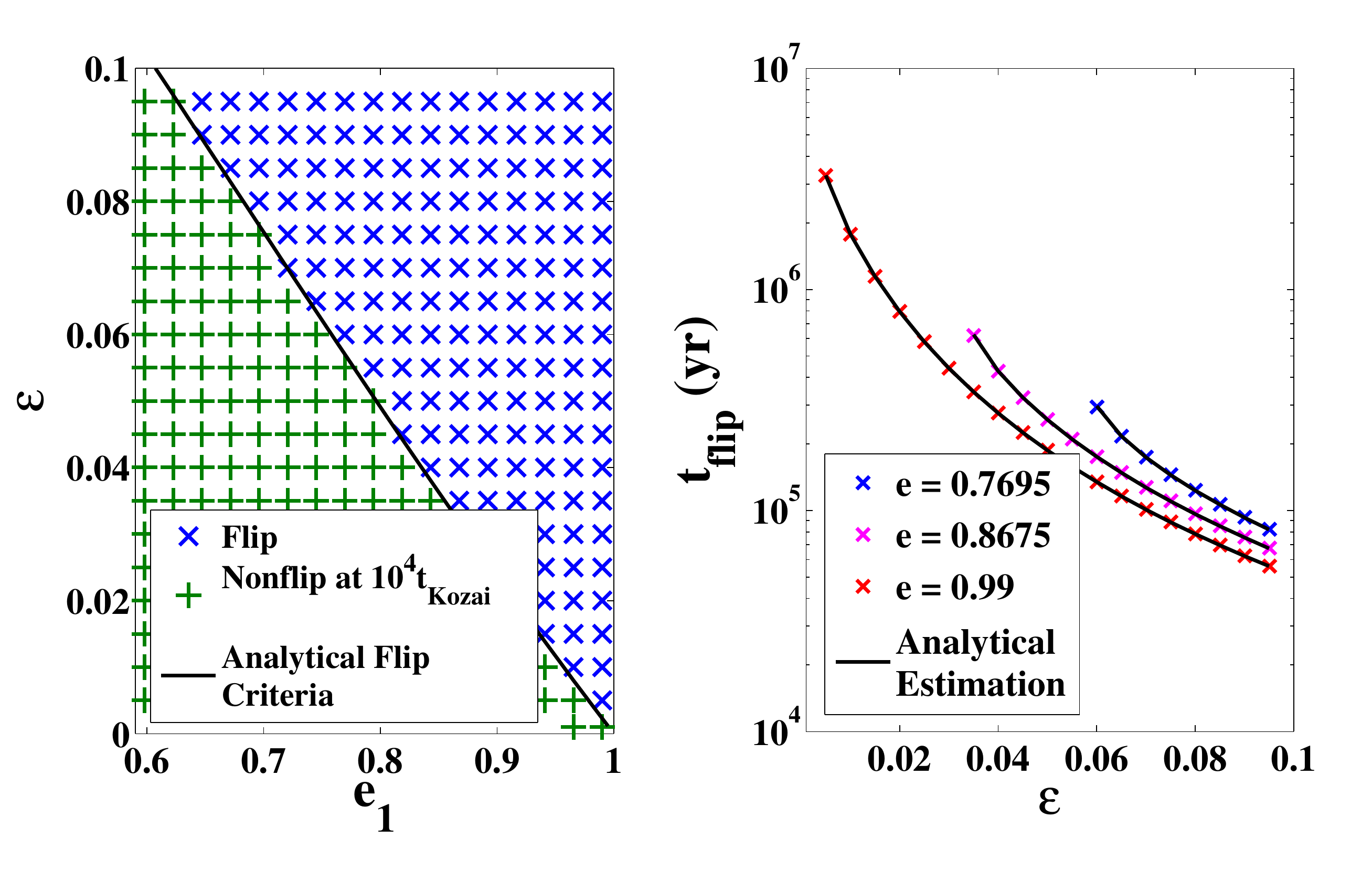}
\caption{{\it Comparisons of the numerical results and the analytic expressions for the point mass dynamical evolutions}. The initial inclination is $i = 5^{\circ}$. Left panel: the numerical results versus the analytic criterion for the flip condition (equation (2)). The black line indicates the analytic criterion. The numerical result is obtained from the secular integration, where the initial condition is: $m_1 = 1M_{\odot}$, $m_2 = 0.1M_{\odot}$, $a_1 = 1$AU, $a_2 = 45.7$AU, $\omega_1 =0^{\circ}$, $\Omega_1=180^{\circ}$. The blue crosses represent the flipped runs and the green pluses represent the runs that do not flip in $10^4$ $t_{Kozai}$, where $t_{Kozai}$ is defined in equation \ref{eqn:S7}. Right panel: the flip timescale for different initial eccentricity. The black line indicates the flip time calculated analytically, and the colored crosses are the flip time recorded in the numerical runs.} \label{fig:flipsum} 
\end{center}
\end{figure}

On the other hand, when $\sin(\varpi) > 1$, $e_1$ decreases first before it increases. Since the flip always occurs at the maximum eccentricity, the flip time is simply:
\begin{equation}
t_{flip} = \int_{e_0}^{e_{min}}\dot{e_1}^{-1}~de + \int_{e_{min}}^{e_{max}}\dot{e_1}^{-1}~de \ .
\end{equation}
We calculate $e_{min}$ with equation (\ref{eqn:S5}) by setting $\cos(\varpi) = 1$ and estimate the flip time. As shown in Figure \ref{fig:flipsum} the analytical flipping time, $t_{flip}$, agrees well with the numerical results. 
 
It is straightforward now to derive the flip condition. Rearranging equation (\ref{eqn:S5}), we find   \begin{equation} 
\epsilon \cos(\varpi_1) = \frac{8e_1^2 - C}{e_1(20+15e_1^2)} \ , 
\end{equation} 
where $C$ is the integration constant (energy at $i=0$) introduced in equation (\ref{eqn:S5}). The difference on left hand side between the initial time and the flip time bound by $\epsilon(1-\cos(\varpi_1))$. When the orbit flips, $e_1\to1$ and the difference on the right hand side is 
\begin{equation}
\frac{8-C}{35}-\frac{8e_1^2-C}{e_1(20+15e_1^2)}  \ . 
\end{equation} 

Thus, a flip will happen when the following condition holds: 
\begin{equation}
\epsilon(1-\cos(\varpi_1))>\frac{8-C}{35}-\frac{8e_1^2-C}{e_1(20+15e_1^2)} \ . 
\end{equation}

Substituting $C$ from the initial condition, we obtain the flip criterion:
\begin{equation}
\epsilon > \frac{8}{5}\frac{1-e_1^2}{7-e_1(4+3e_1^2)\cos(\omega_1+\Omega_1)} \ .
\label{eqn:criterion}
\end{equation}

Figure \ref{fig:flipsum} compares the analytical and the numerical results. The left panel focuses on the flip criterion, whereas the black line represents the analytical criterion, the green plus symbols represent the numerical runs that do not flip in $10^4 t_{Kozai}$, and the blue cross symbols represent the numerical runs that flip. The timescale $t_{Kozai}$ is defined as:
\begin{align}
\label{eqn:S7}
t_{Kozai} = \frac{m_1}{m_2}\Big(\frac{a_2}{a_1}\Big)^3(1-e_2^2)^{3/2}(1-e_1^2)^{1/2}P_{in},
\end{align}
where $P_{in}$ is the period of the inner orbit \citep{Innanen97, Valtonen05}. We start the runs for different eccentricities and inclinations. The analytical criterion agrees well with the numerical results. In the right panel of Figure \ref{fig:flipsum}, we compare the flip timescale for three arbitrarily chosen eccentricities. The analytical results also agree well with the numerical results. Note that the small inclination assumption holds for most of the evolution, as the actual flip has a much shorter duration than the eccentricity growth that precedes the flip.

\section{Systematic Study of $180^{\circ}$ Flips}
We explored the entire $e_1$ and $i_0$ parameters space that can produce flips. We scanned systematically the parameter space of the initial conditions $e_1$, $i$ and $a_2$ and integrate for the secular evolution of the inner orbit in the test-particle limit.  For systems that flipped within 1000 $t_{Kozai}$ we recorded the time when the flip happens, where $t_{Kozai}$ is defined in equation (\ref{eqn:S7}).

\begin{figure}[h]
\begin{center}
\includegraphics[width=1.45in, height=2in]{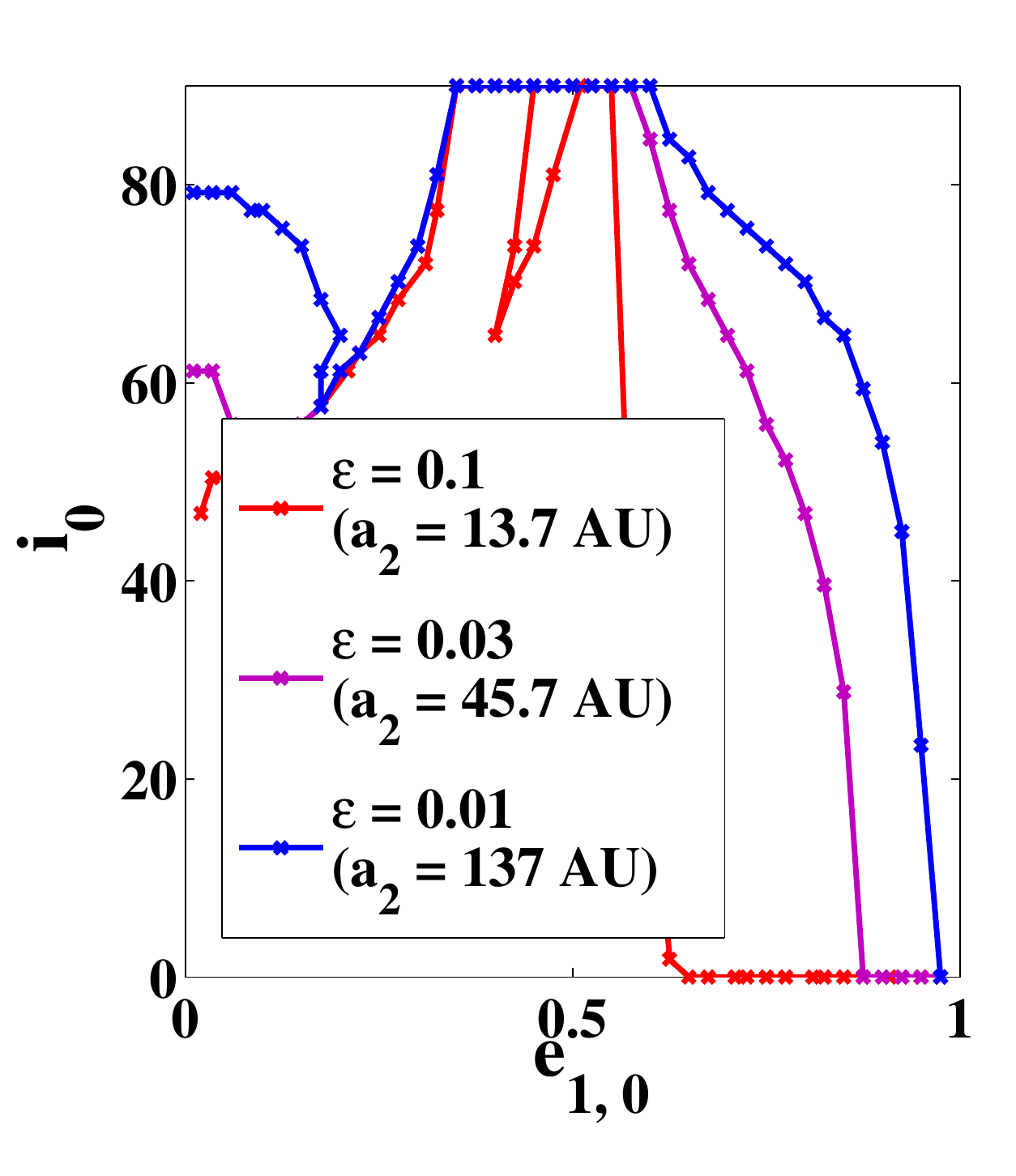}
\includegraphics[width=1.45in, height=2in]{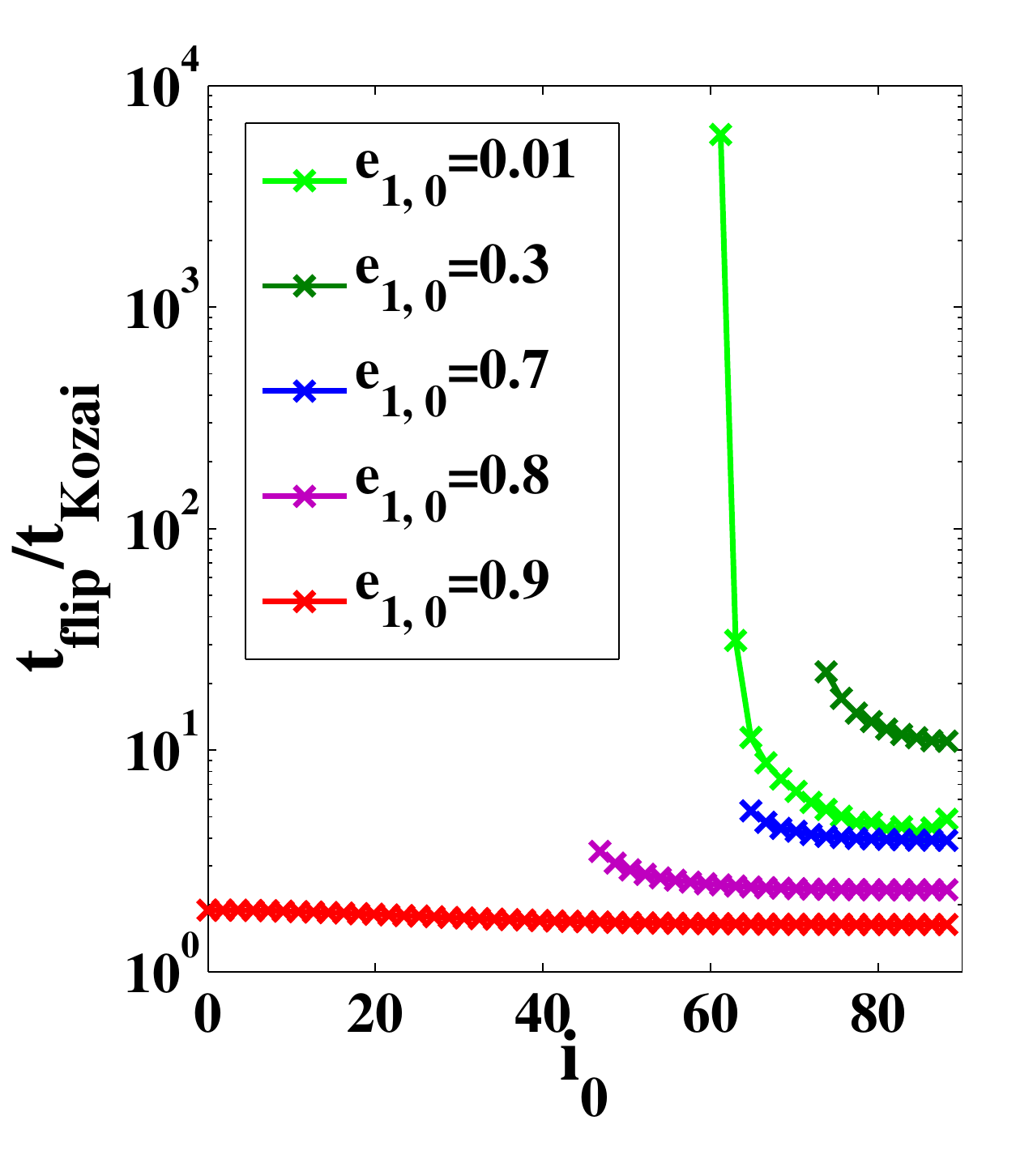}
\caption{{\it The flip condition and the flip time}. Left panel: The flip condition for the whole parameter space of initial $e_1$ and $i$ for three different outer semi-major axes, $a_2$. The initial condition for all the simulations are: $m_1=1M_{\odot}$, $m_2=0.1M_{\odot}$, $a_1=1$ AU, $\omega_1 = 0^{\circ}$, $\Omega_1=180^{\circ}$. $a_2$, $e_1$ and $i$ are different for the runs. The simulations do not include the influence of tides. Initial conditions above the colored lines in the $e_1-i$ plane exhibit an orbital flip. The red line represents the case when $a_2=13.7$ AU ($\epsilon = 0.1$), the purple line represents the case when $a_2=45.7$ AU ($\epsilon = 0.03$) and the blue line represents the case when $a_2=137.5$ AU ($\epsilon = 0.01$). The flip condition agrees well with our analytic estimates for the eccentric coplanar cases. The flip condition is more complicated at moderate $e_1$. Right panel: The flip time for $a_2=45.7$ AU. The flip time is shorter for the HeLi case. Note: when $e_1$ is higher, $t_{Kozai}$ is shorter (see equation \ref{eqn:S7}). Thus, the eccentric coplanar flip time is much shorter than the standard Kozai.} \label{fig:S5} 
\end{center}
\end{figure}

At low eccentricity, the critical inclination (above which the orbit flips) increases. This is consistent with the flip condition of the HiLe mechanism \citep{Lithwick11, Katz11}, where here we have extended Figure 8 of \citet{Lithwick11} to larger initial $e_1$. However, unlike \citet{Lithwick11} that scan the $e_1(\omega=0)$ (i.e., the  minimal eccentricity) and $i(\omega=0)$, we determine the initial conditions that will lead to a flip. For the HeLi case, the result is also consistent with the analytical flip condition described in the \textsection 2. At moderate eccentricity, the behavior of the inner orbit is more complicated, and cannot be easily decried analytically. Figure \ref{fig:S5} depicts the numerical results of the systematic exploration of the parameter space. The left panel of Figure \ref{fig:S5} shows the flip condition for different initial inclinations and  eccentricities, as a function of different $\epsilon$. Not surprisingly, stronger perturbations (i.e., larger $\epsilon$) can cause flips in larger regions of the parameter space. Consistent with \citet{Lithwick11}, we also find that the intermediate regime of $e_{1,0}\sim 0.4$ allows for flips. 

The right panel of Figure \ref{fig:S5} shows the flip time (similar to the right panel of Figure \ref{fig:flipsum}, but this time for different initial inclinations). We normalized the time by $t_{Kozai}$. Note that the flip time of the eccentric coplanar scenario is shorter than that of the HiLe mechanism (as also apparent in the example in Figure 2). In addition, when $e_1 > 0.5$, the flip time is shorter as $e_1$ increases.

\section{Application to Exoplanets and Tidal Disruption Events}
The effect we discovered may have different interesting applications. We briefly mention two of them hereafter. As shown in Figure 2, during the evolution the eccentricity can reach very large values, which can result in a small pericenter distance and collisions between the inner two objects. In addition, if the objects do not collide, this allows for tidal dissipation to take place. Specifically, it shrinks and circularizes the orbit. If tide takes place after the orbit rolls over, a counter--orbiting inner orbit can be produced. This configuration is interesting as the inner orbit is almost coplanar with the outer orbit but goes in the opposite direction.

\subsection{Counter Orbiting Hot Jupiters}
Hot Jupiters -- massive extrasolar planets in a very close proximity to their host star ($\sim1-4$ day orbit) -- are observed to exhibit interesting characteristics. The planet's projected orbital orientation ranges from almost perfectly aligned to almost perfectly anti-aligned with respect to the spin of the star \citep{Albrecht12}.  In other words, the sky projected angle between the stellar spin axis and the planetary orbit (the spin-orbit angle, otherwise known as obliquity) is observed to span the full range between $0^{\circ}$ and $180^{\circ}$. 

Formation theories that rely on a planet slowly spiraling in through angular momentum exchange with the protoplanetary disk produce low obliquities (\citet{Lin86}, but see \citet{Thies11, Batygin12}). The highly misaligned configuration poses a unique challenge to planet formation and evolution models. It was suggested that secular perturbations due to a distant object \citep{Fabrycky07, Veras10, Correia11, Naoz11N, Naoz12}, planet-planet scattering \citep{Ford08, Nagasawa11, Chatterjee11, Boley12} and secular chaos excursions \citep{Wu10} can explain large obliquity, but cannot explain  counter-orbiting configurations. Similar results can be achieved if the star and protoplanetary disk are initially in an aligned configuration for a fine tuned initial condition \citep[see][]{Batygin12}. Furthermore, a test particle can be captured in a $2:1$ mean motion resonance and flip by $\sim 180^\circ$ as migration continues \citep{Yu01}, and test particles in a debris disk can be flipped due to the interaction of a closely separated planet \citep{Tamayo13}.

We note that while the EKL mechanism can produce retrograde orbits (both in the inclination and obliquity sense) \citep{Naoz11N, Naoz12, Naoz13}, it cannot produce counter orbiting Hot Jupiters. This is because these studies initialized the inner planet with small eccentricity, which means that the initial inclination needed to produce large eccentricity oscillations is large $\sim 40^\circ - 140^\circ$. Furthermore, these initial conditions results in an inclination which are more likely to be confined in the same regime \citep{Teyssandier13}.  Thus, the final maximum  hot Jupiters obliquity reached in these experiments and others \citep[e.g.][]{Fabrycky07, Naoz12}  is $\sim 150^\circ$. An obliquity of $\sim 180^\circ$ could be attributed to projection effects. 

The coplanar $\sim180^{\circ}$ flip may play an important role in the obliquity evolution of many exoplanetary systems. Coplanar configurations are naturally produced if the planet and the perturbing object ($m_2$, a star or a planet) are formed in the same disk, or if they are captured in the disk due to hydrodynamic drag. Eccentricity may be excited by planet-planet scattering or interactions with the protoplanetary disk \citep{Ford08, Nagasawa11}. In addition, eccentric gas giant exoplanets are observed at distances larger than 0.1AU from their host star \citep{Ford00}. 

During the orbital flip, the orbit becomes radial ($e_1\to1$), which reduces the pericenter distance, and allows tide to operate. Tidal dissipation shrinks the orbit separation and circularizes it \citep{Matsumura10}. If this happens after the orbital plane rolled over, a counter orbiting Hot Jupiter is formed. 

\begin{figure}[h]
\begin{center}
\includegraphics[width=3in, height=2.in]{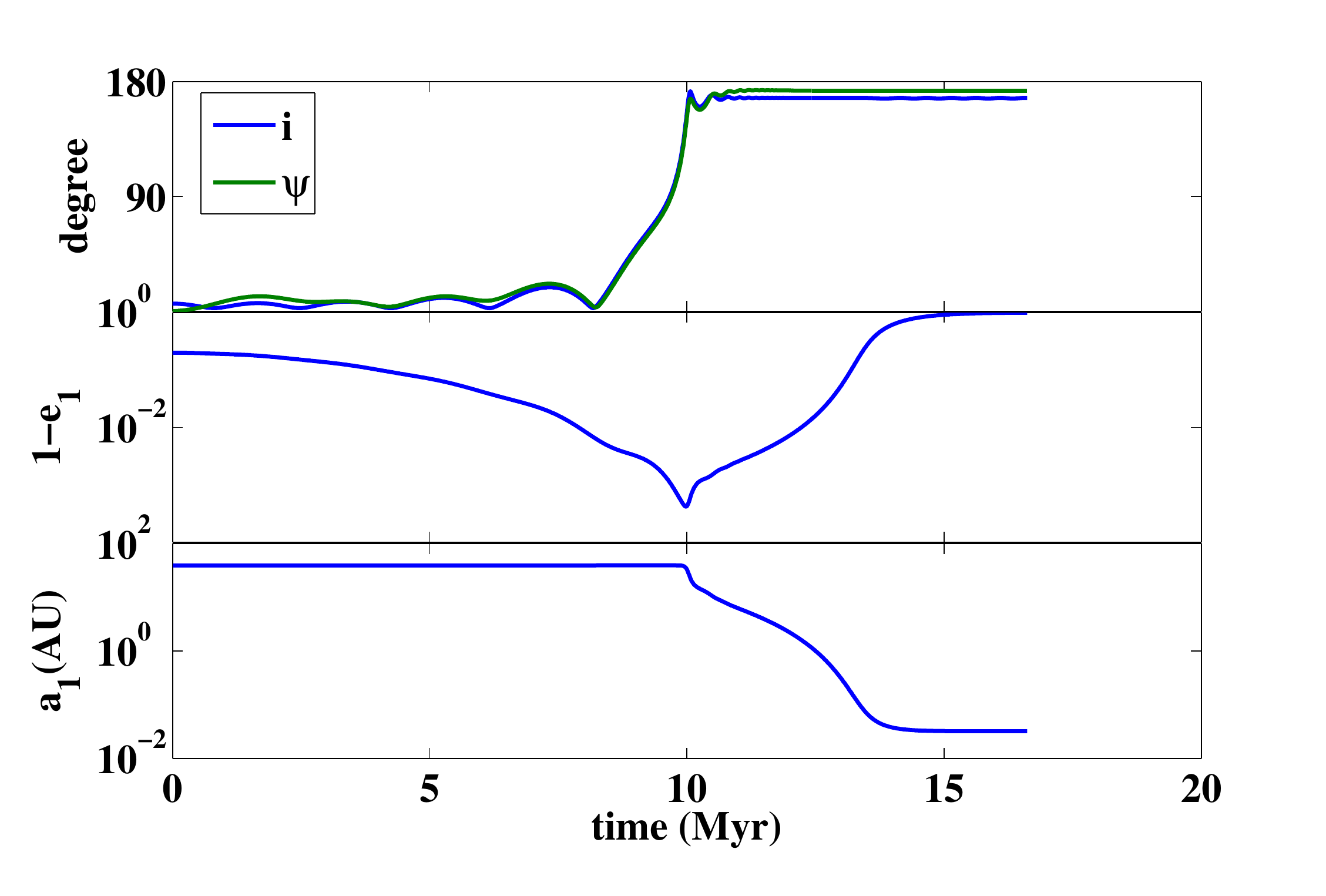}
\caption{{\it The evolution of the inner orbit under gravitational and tidal forces}. The result is obtained by integrating the secular equation of motion. We set the mass and the radius of $m_1$ to be those of the Sun, and the mass and the radius of $m_P$ to be those of Jupiter, and $m_2 = 0.03M_{\odot}$. The initial obliquity angle ($\psi$) is set to be 0. We set $a_1 = 39.35$ AU, $a_2 = 500$ AU, $e_1 = 0.8$, $e_2 = 0.6$, $\omega_1=0^{\circ}$, $\Omega_1 = 180^{\circ}$, $i = 6^{\circ}$ for the initial condition. For tides, we set the dissipation quality factor to be $Q_1 = 10^6$, $Q_J = 10^5$. The orbit flips after $\sim10 $Myrs. During the flip, $e_1 \sim 1$ and the tidal dissipation forces the orbit to decay and circularize. The orbit reaches equilibrium with $\psi \sim 173^{\circ}$, $a_1 \sim 0.032$ AU and $e_1 \sim 0$. General Relativity precession of the inner and outer body is included following \citet{Naoz13GR}.} \label{fig:3} 
\end{center}
\end{figure}

We illustrate this behavior in Figure \ref{fig:3} where the orbit flips within 10Myr from $\sim 6^{\circ}$ to $\sim170^{\circ}$ and the obliquity flips from $0^{\circ}$ to $\sim173^{\circ}$. This orbit reaches its equilibrium state in a circular counter-orbiting configuration with a small semi-major axis (0.032 AU). Such large obliquities may represent the observed retrograde hot Jupiter HAT-P-7 b and HAT-P-14 b, where the sky projected obliquities are $182^{\circ}.5\pm9^{\circ}.4$ and $189^{\circ}.1\pm5^{\circ}.119$, \citep{Winn11}. 

In Figure \ref{fig:3}, we adopt the ``equilibrium tidal" model \citep{Hut81, Eggleton98, Eggleton01}. Its complete set of equations of motion can be found in \citet{Fabrycky07}. Specifically, this approach takes into account the rotation of the star, and the distortion of the planet due to rotation and the tide of the star. In addition, it assumes the viscous timescales of the planet and the star are constant and the tidal quality factor $Q$ is proportional to the orbital period of inner orbit \citep{Hansen10}. In the example we show in Figure \ref{fig:3}, we set the viscous timescale of the star and the planet to 50 years and 0.94 years, respectively, which correspond to the quality factors of $Q\sim10^6$ and $10^5$ for a 10 day orbital period. In this calculation we also include General Relativity precession of the inner and outer body, following \citet{Naoz13GR}.

The example shown in Figure \ref{fig:3} predicts that this counter-orbiting planet has an eccentric coplanar companion. We stress that this does not mean that one should expect a high abundance of counter orbiting planets, nor that even one exists. This mechanism can produce a large range of final inclinations depending on when tides start to dominate. The pericenter distance shrinks before and during the flips, and when tides become important their effect may effectively halt the orbital flip. In addition, this mechanism drives the inner orbit eccentricity to extremely high values and might result in the planet colliding with or tidally disrupted by the star. Calculating the fraction of systems that will result in a counter orbiting planet and the fraction of planets that will collide with the star is beyond the scope of this paper.

\begin{figure}[h]
\begin{center}
\includegraphics[width=3in, height=2in]{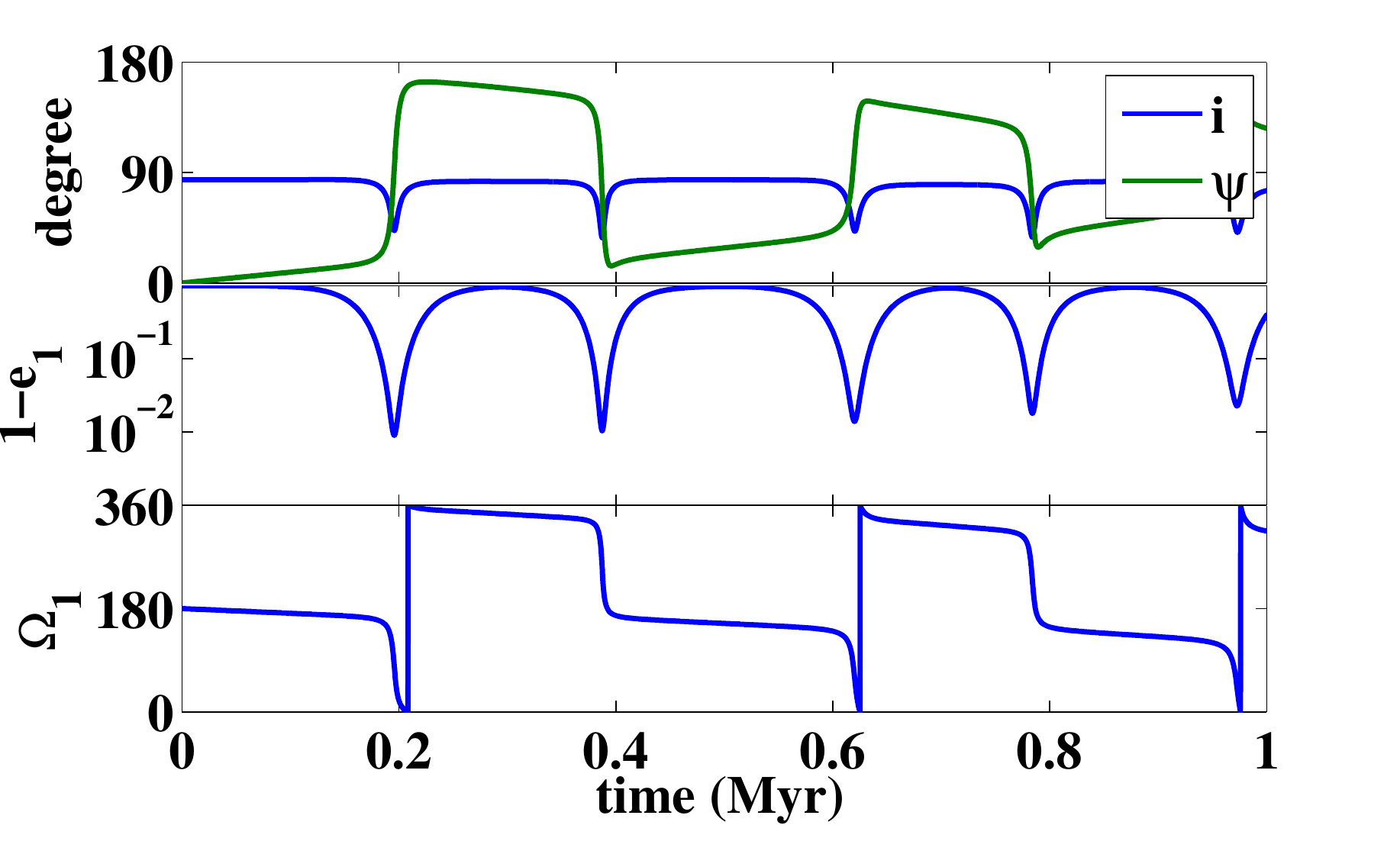}
\caption{{\it The $\sim180^{\circ}$ flip of the spin-orbit angle when the mutual orbital inclination is slightly less than $90^{\circ}$}. We set the mass and the radius of $m_1$ to be those of the Sun, and the mass and the radius of $m_P$ to be those of Jupiter, and $m_2 = 0.03M_{\odot}$. The initial spin-orbit angle ($\psi$) is set to be 0. We set $a_1 = 40$ AU, $a_2 = 500$ AU, $e_1 = 0.01$, $e_2 = 0.6$, $\omega_1=0^{\circ}$, $\Omega_1 = 180^{\circ}$, $i = 85^{\circ}$ for the initial condition.  The top panel shows the point mass dynamical evolution of the inclination and the spin orbit angle, and we can see that during each Kozai cycle and the inclination oscillates, the spin orbit angle flips. In the middle panel, $e_1$ is plotted as a function of time. In the bottom panel, we show that the longitude of the ascending node shifts by $\sim180^{\circ}$ abruptly at the end of each Kozai cycle. This indicates the rapid $\sim180^{\circ}$ flip of the orbit in the x-y plane.} \label{fig:S2} 
\end{center}
\end{figure}

Related to the coplanar flips, we explain the behavior found by \citet{Fabrycky07}, where the spin orbit angle flips in the test particle quadruple limit while the inclination does not flip. In this limit, one of the members of the inner orbit is a test particle and the outer orbit is circular, the z component of the angular momentum is conserved. If the orbit starts prograde $i<90^\circ$ is will remain prograde. However, the obliquity can flip from prograde to retrograde, as shown in the top panel of Figure  \ref{fig:S2}. This is a different kind of flip because the flips occur in the x-y plane (as discussed below). 

In the limit at $i\sim90^{\circ}$, $dJ_1/dt$ is in the direction of $J_1$ and $\Omega_1$ shifts by $180^{\circ}$ \citep{Katz11}. Thus, $J_1$ moves in a straight line across the origin in the x-y plane and the orbit flips by $180^{\circ}$ in the x-y plane. The orbital direction of the inner planet is reversed while the mutual inclination remains less than $90^{\circ}$. 

This can be seen in the movies as well. The flip timescale is the quadrupole Kozai timescale. Because the flip of the orbit is abrupt, tides from the planet cannot respond fast enough to realign the stellar spin to the angular momentum of the inner orbit. As a consequence, the spin-orbit angle crosses $90^\circ$ (Figure \ref{fig:S2}). The behavior also persists when the inclination is less than $90^{\circ}$, but in that case the shift of the longitude of ascending node and the change in obliquity are less than $180^{\circ}$. 

Similar to the HeLi flip, the flip in the x-y plane can also produce $\sim180^{\circ}$ counter-orbiting planets with respect to the stellar spin, however, this requires the perturber's orbit to be nearly perpendicular to the inner orbit. The flip in the x-y plane may also be relevant for gravitational waves emitted by compact object binaries, where the orbital flip changes the polarization angle of the signal. 

% We predict that counter-orbiting planets have eccentric coplanar companions (either a planet or a star). Transit timing variation, direct imaging and radial velocity measurements may find observational evidence for the inner planet and the massive outer object.
%%%%%%%%%%%%%%% HERE %%%%%%%%%%%%%%%

\subsection{Tidal Disruption Events - Systematic Study} 
As mentioned above, the eccentric Kozai--Lidov mechanism (large and small inclination) drives the inner orbit eccentricity to very large values. This reduces the pericenter distance. When an object moves close to $m_1$, the tidal force of $m_1$ can get stronger than the object's self-gravity and hence tidally disrupt the object. For instance, stars may be tidally disrupted by supermassive black holes if they pass very close to the black holes. Tidal disruption of stars by black holes may produce luminous electromagnetic transients that have been observed \citep[e.g.][]{Bade96, Komossa99, Gezari03, Gezari06, Gezari08, Gezari09, van11, Cenko12, Gezari12}. 

\begin{figure}[h]
\begin{center}
\includegraphics[width=3in, height=2.in]{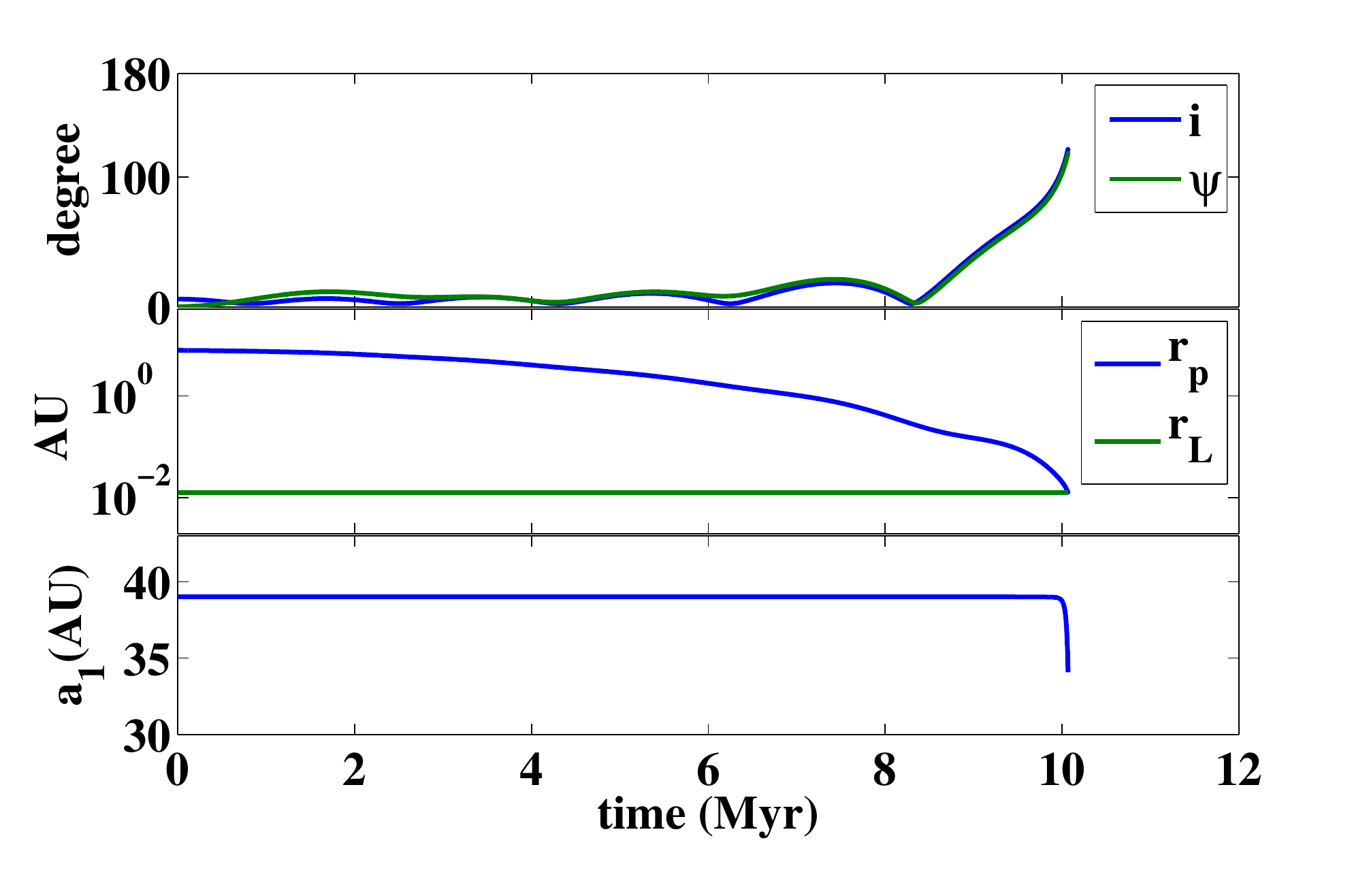}
\caption{{\it An example illustrating a tidally disruption event}. The initial condition is the same as in Figure \ref{fig:3}, except $a_1=39$ AU. Similar to Figure \ref{fig:3}, both tidal dissipation and General Relativity precession effects are included (see text). During the flip, $e_1 \sim 1$ and the tidal dissipation forces the orbit to decay (as shown in the bottom panel). However, the tidal circularization is outran by the eccentricity excitation during the flip, and the object is disrupted before reaching $180^\circ$ when $r_p < r_L$, where $r_L$ is the Roche limit of the object to $m_1$.} \label{fig:disrupted} 
\end{center}
\end{figure}

We show an example of an object passing the Roche limit in Figure \ref{fig:disrupted}. To mimick the case that produces a counter-orbiting exoplanet (e.g. Figure \ref{fig:3}), we use the same initial parameters but with a different semi major axis ($a_1 = 39$ AU). In addition, this calculation includes both tidal dissipation and General Relativity precession effects, similar to Figure \ref{fig:3}. In this case, during the flip, the eccentricity increases, causes the pericenter to reach the Roche limit of the planet and disrupting the planet. 

A very large eccentricity does not immediately imply a tidal dissipation event, since this depends on the initial separation of the orbit. We map the maximum eccentricity that can be reached during the evolution, which may then be useful to examine the likelihood of tidal disruption for specific systems.

\begin{figure}[h]
\begin{center}
\includegraphics[width=3in, height=2.5in]{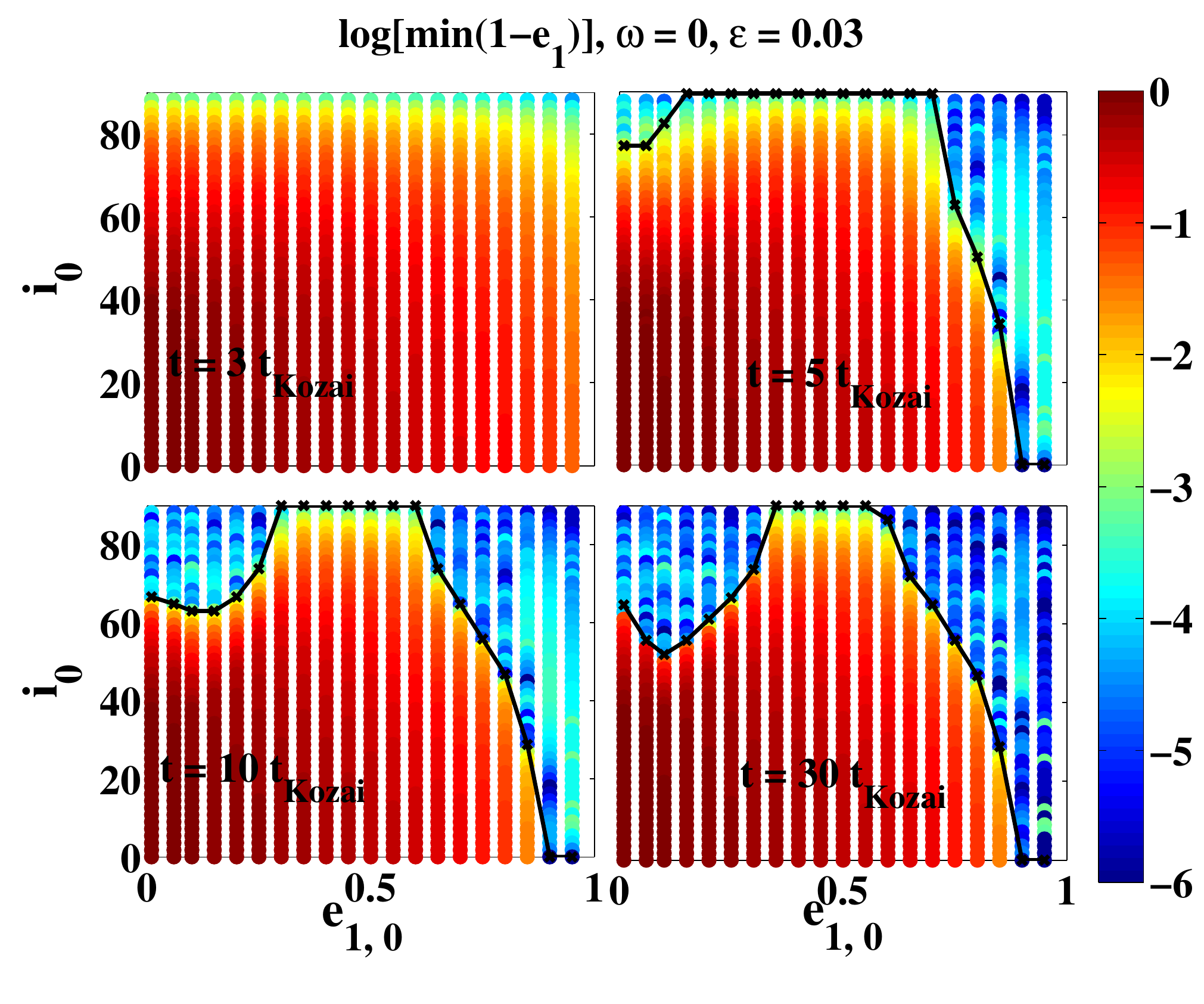}
\caption{{\it The maximum eccentricity}. The maximum eccentricity reached during the secular evolution in time $3 t_{Kozai}$  (upper left panel), $5 t_{Kozai}$ (upper right panel), $10 t_{Kozai}$ (lower left panel) and $30 t_{Kozai}$ (lower right panel) as a function of the initial eccentricity (horizontal axis) and inclination (vertical axis). Tides are not included in the simulation. The initial condition of the runs are $m_1=1M_{\odot}$, $m_2=0.1M_{\odot}$, $a_1 = 1$ AU, $a_2 = 45.7$ AU, $e_2=0.7$, $\omega_1 = 0^{\circ}$, $\Omega_1=180^{\circ}$. The typical eccentricity reached at the first flip is $\sim1-10^{-4}$, and the eccentricity may increase to $\sim1-10^{-6}$ after several flips. The HiLe case reaches the maximum eccentricity later than the LiHe case. The inner orbit flips above the black solid lines.} \label{fig:S6} 
\end{center}
\end{figure}

Specifically, we study the maximum eccentricity reached during the evolution for $\epsilon=0.03$. Since this depends on the time the integration stops, we record the respective maximum eccentricity of the inner orbit for integration times $3 t_{Kozai}$, $5 t_{Kozai}$, $10 t_{Kozai}$ and 30 $t_{Kozai}$. As shown in Figure \ref{fig:S6} the eccentricity of the inner orbit can be very close to one, with  $1-e_{1,max}\sim 10^{-4}$ during the first flip, and $10^{-6}$ over longer time periods. 

This process is relevant for estimating the rates of planet-star collisions \citep{Hellier09, Bear11}, stellar tidal disruptions due to black hole binaries (\citealt{Ivanov05, Colpi09, Chen11, Wegg11, Bode13, Stone12}; Li et al. in prep.), Type 1a supernovae \citep{Katz12}, star-star collisions (e.g. \citet{Perets09F, Thompson11, Katz12, Shappee13, Naoz13, Naoz14}) and gravitational wave sources \citep{OLeary09, Kocsis12}.

\section{Conclusion}
We have presented a new mechanism that flips an eccentric inner orbit by $180^\circ$ starting with a near-coplanar configuration in a hierarchical three body system with an eccentric outer perturber. We use the secular approximation to study the dynamics, and show the agreement between the secular treatment and the N-body simulation in Figure \ref{fig:directInt}. 

The HeLi (high eccentricity low inclination) flip is a different mechanism from the LeHi flip discussed by \citet{Naoz11N, Naoz13}. The underlying resonances causing the large oscillation in the inclination and the flip are different: the LeHi flip is caused by both the quadrupole and the octupole interactions. However,  in the HeLi case, only octupole resonances are in play (see for further discussion in Li et al. in prep). Moreover, for the low inclination case, the orbital evolution is regular, which admits a simple analytic flip criterion and timescale (which were shown to agree with the numerical results in Figure \ref{fig:flipsum}). Specifically, the flip criterion is shown in equation (\ref{eqn:criterion}). In addition, the difference can be seen through the evolution of the orbit: the eccentricity increases monotonically and the inclination remains low before the flip, and the flip timescale of the coplanar case is shorter comparing with the high inclination case (see Figure \ref{fig:2ab} and movies. Finally, we explored the entire $e_1$ and $i_0$ parameters space including both the high inclination and low inclination flips. We studied the flip condition for the initial condition in Figure \ref{fig:S5}. The evolution of the near-coplanar systems is distinct from the exact coplanar systems, since in the exact coplanar systems the net force normal to the orbital plane is zero and thus the orbit cannot flip. Therefore, N-body simulations that assume exactly zero inclination may miss some of the dynamical behavior arises even for small deviations from coplanarity.

Observations of the sky-projected obliquity angle of Hot Jupiters shows that their orbital orientation ranges from almost perfectly aligned to almost perfectly anti-aligned with respect to the spin of the star \citep{Albrecht12}. We showed in the hierarchal, nearly coplanar, three body framework, an initial eccentric inner orbit can flip its orientation by almost $180^\circ$ in the presence of an eccentric companion (Figures 5 and 6). During the planet's evolution its eccentricity is increased monotonically, and thus tides are able to shrink and circularize the orbit. If the planet has flipped by $\sim 180^\circ$ before tidal evolution dominates, a counter orbiting close-in planet can be formed. 

Figure \ref{fig:3} demonstrated this behavior. Not only does the final planet inclination reach $180^\circ$ with respect to the total angular momentum, but also the obliquity. This is because the timescale to torque the spin of the star is much longer than the orbital flip timescale, the spin-orbit angle is similar to the inclination at $\sim 180^\circ$. Therefore, starting with an initially aligned spin orbit configuration, the mechanism presented here can produce counter orbiting close-in planets for a nearly coplanar system. The counter orbiting exoplanets with a $180^\circ$ obliquity angle can be verified using the measured spin-orbit angle. The true spin-orbit angle can be obtained from the sky projected spin-orbit measurement using the Rossiter-McLaughlin method and the line of sight spin-orbit angle measurement using astroseismology.

We note that we do not expect an excess of counter orbiting planets, because this mechanism can drive the inner orbit to an extremely large  eccentricity (see Figure 10) therefore the planet may often end up plunging into the star before circularizing due to tidal effects. A systematic survey of the likelihood of creating counter orbiting planets is beyond the scope of this paper. 

In addition to exo-planetary systems, this mechanism can be applied to many different astrophysical settings, which can tap into the parameter space of hierarchical three body system that has large initial eccentricities and low inclinations. As the eccentricity can be excited to $\sim 1-10^{-6}$ (Figure \ref{fig:S6}), this mechanism may result in an enhanced rate of collisions or tidal disruption events for planets, stars and compact objects with hierarchical three body configuration.

\section*{Acknowledgments}
We thank Konstantin Batygin, Matt Holman, Josh Winn and Boaz Katz for useful remarks. 
 SN is  supported by NASA through a Einstein Post--doctoral Fellowship awarded by the Chandra X-ray Center, which is operated by the Smithsonian Astrophysical Observatory for NASA under contract PF2-130096. BK was supported in part by the W.M. Keck Foundation Fund of the Institute for Advanced Study and NASA grant NNX11AF29G. AL was supported in part by NSF grant AST-1312034.




\bibliographystyle{hapj}
\bibliography{msref}

\end{document}